\documentclass{PoS}
\usepackage{multirow}

\def\n{{\bf \hat n}}

\title{Cross correlation surveys with the Square Kilometre Array}

\ShortTitle{Cross correlation surveys with the SKA}

\author{Donnacha Kirk$^1$, Aur{\'e}lien Benoit-L{\'e}vy$^{1}$, \speaker{Filipe B. Abdalla}$^{1,2}$, Philip Bull$^3$, Benjamin Joachimi$^1$\\

$^1$Department of Physics and Astronomy, University College London,
London WC1E
6BT, UK\\
$^2$Department of Physics and Electronics, Rhodes University, PO Box 94, Grahamstown, 6140 South Africa; \\
$^3$ Institue of Theoretical Astrophysics, University of Oslo, P.O. Box 1029
Blindern, N-0315 Oslo, Norway
\\
E-mail:\email{\\drgk@star.ucl.ac.uk\\benoitl@star.ucl.ac.uk\\fba@star.ucl.ac.uk}

}

\abstract{By the time that the first phase of the Square Kilometre Array is deployed it will be able to perform state of the art Large Scale Structure (LSS) as well as Weak Gravitational Lensing (WGL) measurements of the distribution of matter in the Universe. In this chapter we concentrate on the synergies that result from cross-correlating these different SKA data products as well as external correlation with the weak lensing measurements available from CMB missions. We show that the Dark Energy figures of merit obtained individually from WGL/LSS measurements and their independent combination is significantly increased when their full cross-correlations are taken into account. This is due to the increased knowledge of galaxy bias as a function of redshift as well as the extra information from the different cosmological dependences of the cross-correlations. We show that the cross-correlation between a spectroscopic LSS sample and a weak lensing sample with photometric redshifts can calibrate these same photometric redshifts, and their scatter, to high accuracy by modelling them as nuisance parameters and fitting them simultaneously cosmology. Finally we show that Modified Gravity parameters are greatly constrained by this cross-correlations because weak lensing and redshift space distortions (from the LSS survey) break strong degeneracies in common parameterisations of modified gravity.}

\FullConference{
Advancing Astrophysics with the Square Kilometre Array\\
June 8-13, 2014\\
Giardini Naxos, Italy}

\begin{document}

\section{Introduction}

The Square Kilometre Array is a facility which will be able to provide huge advances in several areas of astronomy and cosmology. It will be able to map the large scale distribution of galaxies as well as measure the weak gravitational lensing signal from far away objects. The information from these probes is not fully independent as both the LSS and WGL signals depend on the large scale distribution of matter in our Universe. 

The combination of the information encoded in the large scale distribution of galaxies and in the signal from weak gravitational lensing can significantly increase our knowledge of the Universe we live in. Weak Gravitational Lensing (WGL) does not respond to the bias of galaxies and hence the combination of the signal from WGL and LSS can help constrain the bias of galaxies tracing the Large Scale Structure. Furthermore WGL usually requires photometric redshifts to yield an estimate of the redshift distribution of such galaxies. The cross correlation between WGL samples with a spectroscopic redshifts samples can help calibrate the redshift distribution of WGL galaxies. Finally if we consider theories of modified gravity, WGL combined with the Redshift Space Distortion (RSD) signal in the LSS survey can together probe potential modifications to the Poisson equation as well as the equation which determines how light bends in the presence of matter in separate ways. This breaks a very powerful degeneracy in the common parameterisations of deviations from GR, hence their joint constraints are orders of magnitude more powerful than individual constraints.

In this chapter we take surveys equivalent to those possible with SKA phase 1 as well as SKA phase 2 and estimate how well the combination of probes can calibrate photometric redshifts and improve the figures of merit for Dark Energy and Modified Gravity. We consider several scenarios, notably we describe as SKA 1 early scenarios which are surveys which will be available during a construction phase of the SKA before phase 1 is completed. We assume that there is both a continuum and a line survey for the Weak Lensing and the Large Scale Structure respectively. The Continuum survey would have longer baselines and would be focused on weak lensing and is similar to the survey outlines in the Weak Lensing chapter in this science book. The redshifts for these galaxies would come from matching to optical galaxies with photometric redshifts. The LSS surveys are assumed to be galaxies found in line emission mainly from the signal present in the core of the SKA. We make little distinction between the technologies needed for such surveys but note that a proper UV distribution is needed for a Weak lensing survey and some necessary sensitivity is needed for the line survey. These SKA1 early surveys are effectively 1000 and 5000 sq degree surveys assuming two thirds of the SKA phase 1 sensitivity. We also assume two SKA phase 1 surveys which would reach signals of around 100 $\mu$Jy in the case of a line survey. The Weak lensing survey is assumed not to be larger than 5000 sq degrees in the case of phase 1 as anything more would be infeasible given the sensitivity of the instrument. We assume a larger LSS survey of 30000 sq degs with phase 1 obtaining galaxies and redshifts. For phase 2 we assume both WGL and LSS surveys covering the same amount of the sky with 5000 and 30000 square degrees respectively where increased area trades off with decreased depth. 

These surveys are summarised in Table 1 and 2 and their respective redshift distribution is the same assumed in other chapters of this science book.

\section{Formalism for Cross-Correlations}

In the following we will provide forecasts for various cross-correlations of cosmological probes using the Fisher Matrix formalism (see Heavens (2009) for an overview). The Fisher matrix is defined as
\begin{equation}
F_{\alpha\beta} \equiv \left< H_{\alpha\beta} \right> = \left< -\frac{\partial^{2} \ln L}{\partial p_{\alpha}\partial p_{\beta}} \right>\;,
\end{equation}
where $L$ is the likelihood and where $p_\alpha$ are cosmological parameters. Assuming a Gaussian likelihood for the data and a cosmology-independent data covariance, the Fisher matrix is given by Tegmark, Taylor \& Heavens (1997)
\begin{equation}
\label{fm}
F_{\alpha\beta} = \sum^{\ell_{max}}_{\ell=\ell_{min}} \sum_{(i,j),(m,n)} \frac{\partial D^{(ij)}(\ell)}{\partial p_{\alpha} }\;{\rm Cov}^{-1} \left[ D^{(ij)}(\ell),D^{(mn)}(\ell) \right] \frac{\partial D^{(mn)}(\ell)}{\partial p_{\beta} }\;.
\end{equation}
For any unbiased estimator the Fisher matrix provides a lower bound on the marginalised error of a parameter $p_\alpha$, via the Cramer-Rao inequality, $\Delta p_{\alpha} \geq \sqrt{(F^{-1})_{\alpha\alpha}}$. The data vector, $D^{(ij)}(\ell)$, consists of angular power spectra, $C_{XY}^{(ij)}(\ell)$ as a function of multipole $\ell$, for a given combination of probes, $X$ and $Y$, and a pair of redshift bins $i$ and $j$. The covariance matrix in Eq. (\ref{fm}) is assumed to be Gaussian and takes into account shot noise as well as cosmic variance contributions, see e.g. Takada \& Jain (2004) for the weak lensing case.

Following the notations of Joachimi \& Bridle (2010), the angular power spectra of the cross-correlations between the various quantities related to the gravitational potential or the matter density can be written in a generic way  using the Limber approximation,
\begin{equation}
C^{(ij)}_{XY}(\ell) = \int _0^{\chi_{\rm hor}} d\chi\; \frac{w^{(i)}_X(\chi)\; w^{(j)}_Y(\chi)}{f_K^{2}(\chi)}\; P_\delta(\ell/f_K(\chi), \chi)\;,
\end{equation}
where $\chi$ is comoving distance, and $\chi_{\rm hor}$ the comoving horizon. The matter power spectrum is denoted by $P_\delta$ and the comoving angular diameter distance by $f_K(\chi)$. The kernels for the different cosmological probes are given by the following equations,
\begin{eqnarray}
w^{(i)}_\epsilon(\chi) &=& \frac{3 \Omega_m H_0^2}{2 c^2} \frac{f_K(\chi)}{a(\chi)} \int_\chi^{\chi_{\rm hor}} {\rm d}\chi' p^{(i)}(\chi')\; \frac{f_K(\chi'-\chi)}{f_K(\chi')}   ~~~\mbox{(galaxy weak lensing)}\;;\\ 
w^{(i)}_n(\chi) &=& b_g(\ell/f_K(\chi),\chi)\; p^{(i)}(\chi)   ~~~\mbox{(galaxy clustering)}\;;\\
w_{\rm CMB}(\chi) &=& \frac{3 \Omega_m H_0^2}{2 c^2} \frac{f_K(\chi)}{a(\chi)} \frac{f_K(\chi_*-\chi)}{f_K(\chi_*)}   ~~~\mbox{(CMB lensing)}\;.
\end{eqnarray}
Here, $a$ denotes the scale factor, $p^{(i)}$ the redshift probability distribution for a galaxy sample $i$ (either a broad tomographic bin or a narrow range defined via spectroscopic redshift), and $b_g$ the galaxy bias which can vary as a function of scale and redshift. CMB lensing has a single source distance, $\chi_*$, to the last scattering surface.

Note that the Limber approximation assumes that the kernels involved are broad in the line-of-sight direction, which breaks down in the case of spectroscopic clustering information. In this case we replace the formalism outlined above with the exact calculation as detailed in Padmanabhan et al. (2007).

We forecast constraints for our weak lensing (WGL) surveys alone, $\{C_{\epsilon\epsilon}^{(ij)}(\ell)\}$, large-scale structure (LSS) surveys alone (galaxy clustering including redshift-space distortions), $\{C_{nn}^{(ij)}(\ell)\}$, and the joint WGLxLSS analysis including cross-correlations, $\{C_{\epsilon\epsilon}^{(ij)}(\ell),C_{n\epsilon}^{(ij)}(\ell),C_{nn}^{(ij)}(\ell)\}$. We also provide an assessment of the constraining power of CMB lensing with SKA LSS, $\{C_{{\rm SKA}\,nn}^{(ij)}(\ell), C_{{\rm CMB}\, \epsilon\epsilon}^{(ij)}(\ell)\}$, and CMB lensing with SKA WGL probes, $ \{ C_{{\rm SKA}\,\epsilon\epsilon}^{(ij)}(\ell), C_{{\rm CMB}\, \epsilon\epsilon}^{(ij)}(\ell)\}$.

\section{Weak lensing - galaxy position cross correlations}

The SKA survey will provide very competitive measurements of multiple cosmological probes. Combination of different probes allows the breaking of degeneracies between parameters which provides better control of systematics and better constraints on parameters of cosmological interest than any individual probe can achieve.

In this section we consider the combination of an SKA Weak Lensing (WGL) survey (with photometric redshifts provided externally) and an SKA galaxy position i.e. Large-Scale Structure (LSS) survey with spectroscopic quality redshifts.

\begin{figure}
\includegraphics[width=1.1\columnwidth]{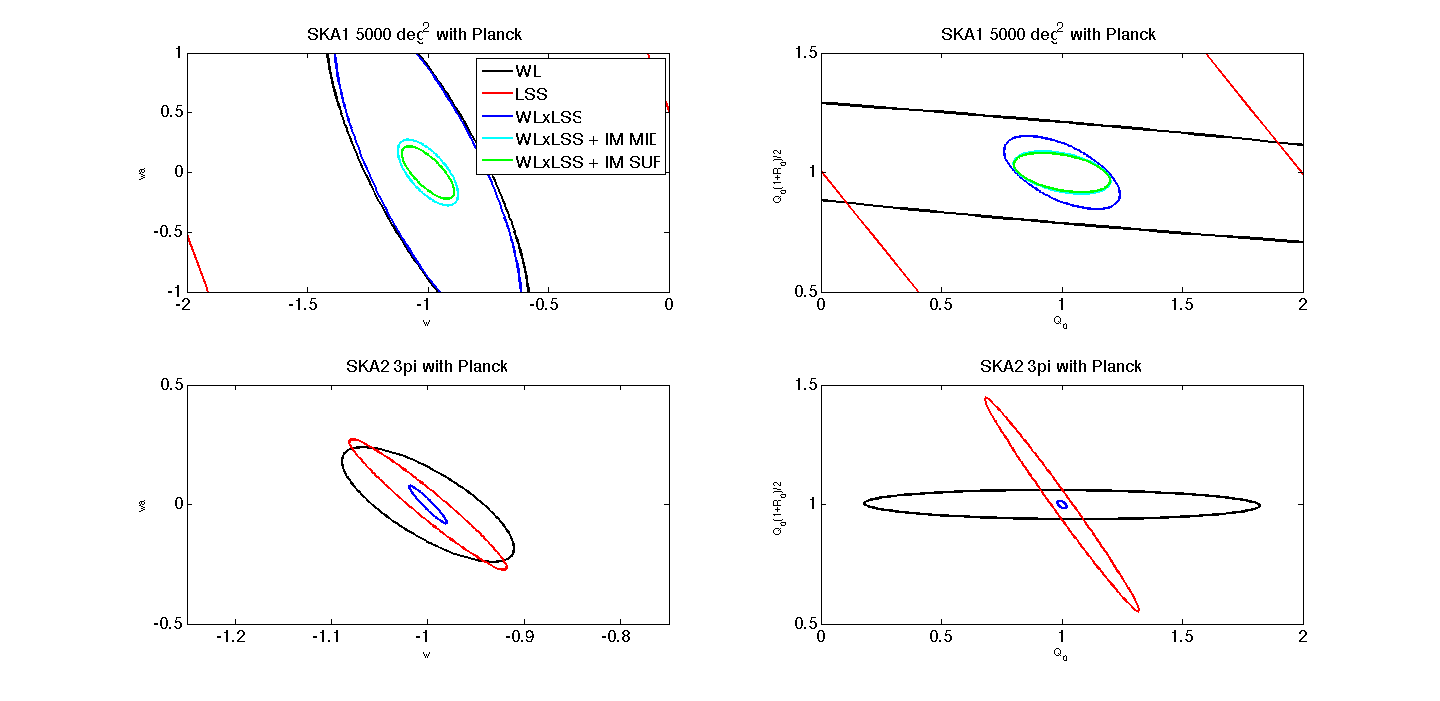}
\caption{Constraints on dark energy [left panels] and deviations from GR [right panels] for SKA1 over 5,000deg$^2$ [top panels] and SKA2 over 30,000deg$^2$ [bottom panels] including Planck priors. Black ellipses show photometric WGL constraints only. Red ellipses show spectroscopic LSS constraints only. Blue ellipses show the combination of WGL and LSS including cross-correlations. Cyan ellipses show this WGLxLSS constraint combined independently with an SKA intensity mapping (IM) survey using the MID instrument and the green contours the same but with the SUR instrument. All constraints are 68\% confidence contours.}
\label{fig:DE_MG_crosscorr}
\end{figure}

Fig. \ref{fig:DE_MG_crosscorr} shows the headline constraints on dark energy (left panel) and deviations from General Relativity (right panel). Table \ref{tab:survey_details} gives details of the assumed area and number density of sources for the different surveys we forecast. 

\begin{table*}
   \centering
   \begin{tabular}{ |l|c|c|c|c|c c c| } 
   \hline
Survey &  \shortstack{Area [deg$^2$] \\ Photo-z} & \shortstack{Area [deg$^2$] \\ Spec-z} &  \shortstack{$n_{g}$  [arcmin$^{-2}]$ \\ Photo-z} &  \shortstack{$n_{g}$ [arcmin$^{-2}]$ \\ Spec-z} & \multicolumn{3}{c}{\shortstack{DE FoM inc. Planck \\ WGL   LSS   WGLxLSS}} \vline \\
\hline
SKA1 early & 1,000     & 1,000	& 4.8 & -   & 3.0 & - & -  \\
SKA1 early & 5,000     & 5,000	& 1.6 & -   & 3.6 & - & -  \\
SKA1       & 5,000     & 5,000	& 4.2 & 1.8 & 16 & 0.05 & 19.2  \\
SKA1       & 5,000     & 30,000	& 4.2 & 4.7 & 16 & 1.8 & 956  \\
SKA2       & 5,000     & 5,000	& 44  & 22  & 182 & 320 & 6362  \\
SKA2       & 30,000    & 30,000	& 20  & 10  & 425 & 1042 & 13150  \\
\hline
   \end{tabular}
   
   \centering
   \begin{tabular}{ |l|c|c|c c c| } 
   \hline
Survey &  \shortstack{Area [deg$^2$] \\ Photo-z} & \shortstack{Area [deg$^2$] \\ Spec-z} & 
\multicolumn{3}{c}{\shortstack{MG FoM inc. Planck \\ WGL   LSS   WGLxLSS}} \vline \\
\hline
SKA1 early & 1,000     & 1,000	&  1.9 & - & -  \\
SKA1 early & 5,000     & 5,000	&  2.0 & - & -  \\
SKA1       & 5,000     & 5,000	&  8.0 & 4.7 & 415  \\
SKA2       & 5,000     & 5,000	&  69 & 257 & 7429  \\
SKA2       & 30,000    & 30,000	&  118 & 583 & 29947  \\
\hline
   \end{tabular}
   \caption{(Top) Summary of SKA survey forecasts and Dark Energy Figure of Merit results including Planck priors.(Bottom) Summary of survey forecasts For the modified gravity Figure of Merit from SKA phases one and two. The MG FOM here is defined in the same way as the DE FOM with the parameters of interest being the constants modifying the relation between the potentials.}
   \label{tab:survey_details}
\end{table*}

The WGL surveys have photo-z quality redshifts. SKA1 early is assumed to get these from a DES-like survey so we assume photo-z error, $\delta_z = 0.07(1+z)$ and 5 tomographic bins of equal number density out to $z=2$. SKA1 and SKA2 are assumed to get redshifts from a Euclid-like survey with $\delta_z = 0.05(1+z)$ and 10 tomographic bins of equal number density out to $z=2$.

The LSS surveys have galaxy redshift distributions described in the HI bias and simulations chapter of this book. Our forecasts assume 20 tomographic bins up to $z=0.6$ for SKA1 and 40 tomographic bins up to $z=2.0$ for SKA2. We use the exact $C(\ell)$ formalism (not the limber approximation) and include the effects of Redshift Space Distortions (RSDs) according to the formalism of Kaiser (1987).  Both these effects are neglected in the WGL forecasts because the broad tomographic bins make their impact negligible. 

Our FM analysis forecasts constraints for a set of cosmological parameters: \\ $\{ \Omega_{m}, \Omega_{b}, \Omega_{DE}, w_{0}, w_{a}, h, \sigma_{8}, n_{s}, b, Q_{0}, Q_{0}(1 + R_{0})/2 \}$. As well as the standard wCDM parameters, $b$ is a free amplitude on galaxy bias for each shell and $Q_{0}, Q_{0}(1 + R_{0})/2$ are parameterisations of deviations to General Relativity that modify the Poisson equation and the ratio of metric potentials, our ability to constrain these parameters quantifies our ability to test gravity on cosmic scales, see Kirk et al. (2013) for more detail. When quoting constraints on dark energy we marginalise over the cosmological parameters and galaxy bias but keep the modified gravity parameters fixed. When quoting constraints on modified gravity we marginalise over cosmology, including $w_{0}$ and $w_{a}$, and galaxy bias. Priors consistent with the latest Planck temperature constraints are included and the same used in the BAO chapter in this science book.

It is worth remarking that our forecasting approach is not the only method that has been used to estimate the joint power of a photometric WGL survey and a spectroscopic LSS survey. We treat all our observables as projected angular power spectra, $C(\ell)$s. The alternative is to model the LSS survey with a full 3D $P(k,z)$ analysis then combine with WGL $C(\ell)$s. Our projection over the line-of-sight width of a tomographic bin throws away information compared to a full 3D analysis. For a photo-z WGL survey the photo-z scatter and the broad lensing kernel combine to produce a redshift resolution floor, below which precision cannot be pushed. This means our relatively coarse redshift binning captures all the available information. The spectroscopic survey has no such limitation, this is why we use a very large number of narrow tomographic bins to capture equivalent redshift-evolution information as contained in a full 3D analysis. The advantage of our approach is that all observables are in the same form and cross-correlations arise naturally. Mixing a 2D and 3D analysis leads to some rather ad-hoc formalisms to combine different observables.

Our results finding significant improvement from cross-correlation agree with Gaztanaga et al. (2013) who use a mixed 2D and 3D formalism. However, both Cai \& Bernstein (2012) and de Putter, Dore \& Takada (2013) adopt a similar mixed approach but see little improvement from cross-correlation. This is an active area of research and discussion on the correct formalism or even how to accurately compare results is continuing.

DE Figure of Merit (FoM, see Albrecht et al. (2006)) for our various survey combinations are shown in Table \ref{tab:survey_details}. The joint WGLxLSS constraints are extremely strong. While the LSS surveys for both SKA1 and SKA2 favour large areas over depth, the joint WGLxLSS constraint for SKA2 prefers the deeper but smaller 5,000deg$^2$ survey over the shallower 30,000 deg$^2$ survey. The Modified Gravity constraint shown in the right panel of Fig. \ref{fig:DE_MG_crosscorr} has a WGLxLSS constraint that is orders of magnitude stronger than either probe alone. This is due to pronounced degeneracy breaking from the combination of a probe using light (WGL, sensitive to the sum of metric potentials $\Psi + \Phi$) and a probe using galaxies as non-relativistic tracers (LSS, sensitive to the Newtonian potential $\Psi$).

SKA1 WL + LSS represents a formidable dataset but it will of course not be the only one available. Dedicated optical surveys such as DES and eBOSS will be available and relatively mature by the completion of SKA1. In terms of number density and area a representative SKA1 survey configuration is comparable to DES+eBOSS with perhaps marginally better photo-z accuracy achieved with SKA1. We find that SKA1 (WGL+LSS+IM) provides comparable constraints on DE to a representative DES+eBOSS forecast while SKA1 significantly outperforms (more than a factor of five improvement) DES+eBOSS in constraining deviations from GR, probably driven by wider sky coverage for the spec-z survey. The fact that both probes are collected by the same instrument offers benefits for the understanding and control of systematics. Of course having datasets of comparable statistical power from both optical and radio surveys is of considerable scientific value of itself. It is with SKA2 that the project becomes definitively world-leading with much higher WL source density than the Euclid mission and a spectroscopic quality LSS survey covering the full area and depth of the WL measurement.

\section {Cross correlations for redshift calibration.}

Photometric redshifts are less accurate but much faster and cheaper to gather than spectroscopic redshifts. SKA will allow a large number of spectroscopic-quality redshifts to be gathered as it measures both galaxy position and redshift via the 21cm line for low redshift galaxies. For higher redshift galaxies used for example in the WGL survey it will require photometric-quality redshifts gathered by other surveys like Euclid. 

A photometric redshift from a Euclid-like experiment can have a random scatter compared to the true redshift of order $\delta_z \sim$ 0.05(1+z). As well as this Gaussian scatter on the true redshift (which can lead to objects being assigned to the incorrect tomographic bin). We note that the true redshift distribution can become much broader than the extent of the corresponding tomographic bin. There are problems estimating the true mean redshift of a certain tomographic bin and what are known as ``catastrophic outliers'', galaxies whose redshift has been severely misidentified due to failures in the photo-z estimation pipeline, usually these are defined as estimates more than $3\sigma$ away from the true redshift, which e.g. can occur at optical wavelengths if the the Lyman and Balmer breaks are confused. Systematic shifts in the redshift distribution of tomographic bins can induce significant biases in cosmological analysis, so that the mean of these distributions needs to be known to better than a few parts in a thousand.

Cross-correlation with a spectroscopic survey which covers some or all of the same galaxy distribution as the photometric survey can be used to identify and mitigate these errors and calibrate the photometric redshift distribution.

We repeat the WGLxLSS forecasts made above but include now a range of nuisance parameters, $\{ \delta_{z,i}, \Delta b_{z,i} \}$, which quantify our uncertainty on the photometric redshifts. Here $\delta_{z,i}$ is the Gaussian uncertainty on the photo-z estimate in each photo-z bin, $i$, with fiducial values 0.05 and $\Delta b_{z,i}$ is the bias on the mean redshift of bin $i$ due to photo-z mis-estimation with mean values 0. We ignore catastrophic outliers in this analysis. Each of these parameters is allowed to vary in our FM analysis and the inclusion of the WGLxLSS cross-correlation allows them to be constrained.

\begin{figure}
\includegraphics[width=1.00\columnwidth]{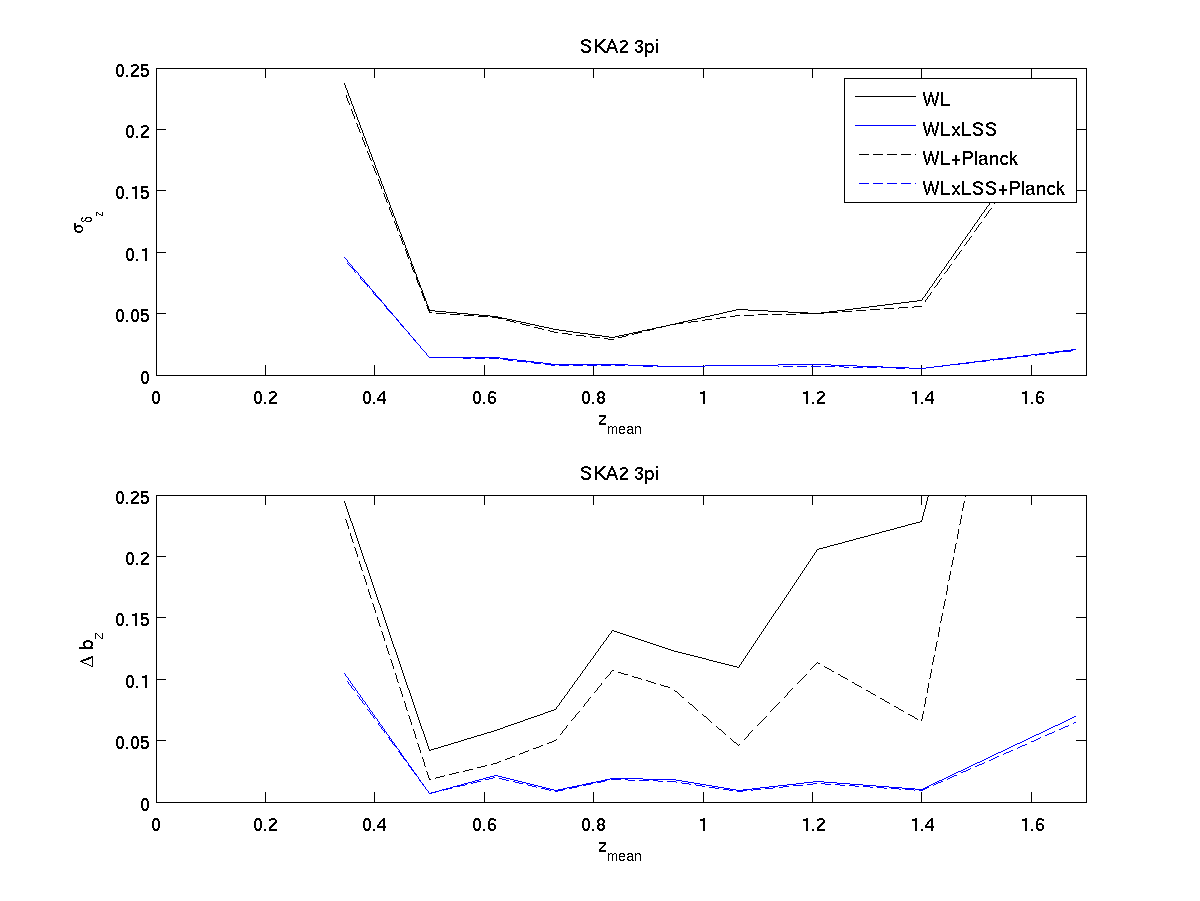}
\caption{Constraints on photometric redshift error [top] and bias on mean redshift [lower] as a function of mean redshift of each tomographic bin for SKA2 over 30,000deg$^2$. WGL- only constraints are shown in black, WGLxLSS constraints in blue. Constraints including Planck priors are shown as dashed lines.}
\label{fig:deltaz_zbias}
\end{figure}

See Fig \ref{fig:deltaz_zbias} for constraints on these nuisance parameters for the SKA2 $3\pi$ survey. Cross-correlation of the WGL survey with the spectroscopic LSS survey improves our estimate of the photo-z distribution by up to  factor of 10. We emphasise that this is a cross-correlation between photometric WGL and spectroscopic LSS, in future we intend to study the cross-correlation between a LSS analysis using the photometric WGL galaxies and the spectroscopic LSS survey. This should prove even more effective at calibrating the photo-z errors because it is not hindered by the broad geometric kernels which restrict redshift resolution in WGL.

\section{Cross-correlations with CMB lensing}

The gravitational potential of the large scale structure generates a deflection of the trajectories of the Cosmic Microwave Background (CMB) photons. This effect, known as CMB lensing  consists of a remapping of the underlying unlensed temperature and polarization fields.
Mathematically, CMB lensing is described as follows:
We introduce a vector field ${\bf d}(\n)$ (the deflection field) such that the lensed temperature $T(\n)$ and
unlensed temperature $\tilde T(\n)$ are related by
\begin{equation}
T(\n) = \tilde T(\n + {\bf d}(\n))
\end{equation}
and analogously for the Stokes parameters $Q(\n)$, $U(\n)$ which describe linear CMB polarization.

To lowest order in perturbation theory, the deflection field ${\bf d}(\n)$ is the gradient of a scalar lensing potential (i.e.~${\bf d}(\n) = \nabla\phi(\n)$)
which can be written as a line-of-sight integral:
\begin{equation}
\phi(\n) = -2 \int d\eta \frac{\chi(\eta-\eta_{\rm{rec}})}{\chi(\eta_{\rm{rec}})\chi(\eta)}\Psi(\chi \n, \eta),
\end{equation}
where $\Psi$ is the Newtonian potential, $\eta$ is conformal time, $\eta_{\rm{rec}}$ is the epoch of last scattering, and  $\chi$ is the angular diameter distance in comoving coordinates, see Lewis \& Challinor (2006) for a review).

CMB lensing modifies the Gaussian structure of the primary anisotropies and generates a correlation between the temperature and its gradient Hu (2000). These couplings can efficiently be used to construct an estimator, quadratic in the observed temperature, that can be applied to data to recover the lensing potential Okamoto \& Hu (2003).

High resolution and high sensitivity CMB experiments can therefore provide a new cosmological probe of the large-scale structure of the Universe that is complementary to that obtained from galaxy surveys. Indeed CMB lensing is mostly sensitive to structure located in the $1 \leq z \leq 5 $ range.

The lensing potential reconstructed from CMB lensing can be thought of as the projection on the sky of all the mass distribution up to the last scattering surface. As such we expect significant angular cross-correlation with the large-scale structure observables of radio surveys. The redshift dependence of the CMB lensing kernel means that, to ensure a strong cross-correlation, we ideally want a LSS survey with many galaxies at $z \gg 1$ and a WGL survey with many galaxies at $Z \gg 2$. SKA will produce very useful surveys in this respect, more useful than Euclid for example when considering this cross-correlation with CMB lensing.

The first reported detection of the gravitational lensing of the Cosmic Microwave Background was made by correlating WMAP data with the radio galaxy counts from the NRAO VLA sky survey (NVSS) (Smith, Zahn \& Dore (2007), Hirata et al. (2007)).
With the advent of arcminute scale CMB experiments (ACT, SPT), and full-sky CMB data from Planck, prospects for cross-correlation between CMB lensing and future radio data are extremely encouraging.

SKA will provide both precise measurement of the position and the shape of galaxies. In the following we use the same redshift distributions assumed in the Weak lensing chapter from this science book, and consider a value constant for the bias $b=1$.

On the CMB lensing side, we consider the following surveys:
\begin{itemize}

\item Planck. We consider the Planck lensing potential from the Planck 2013 results (Planck collaboration (2013)). The current Planck lensing map should be replaced in October 2014 and will include the full Planck data. We thus  multiply the 2013 lensing noise levels by a factor $0.8$ to account for this added data.

\item South Pole Telescope (SPT). The SPT collaboration will observe the CMB polarization anisotropies to arcminute resolution on a 2500 sq. deg. patch in the southern hemisphere. The resulting lensing map will then be of a much better quality than the Planck one, but on a smaller area on the sky. The forecasted noise levels were provided by G.Holder and G. Simard (private communication).

\end{itemize}

In the following we quantify the level of detection of the cross-correlation between CMB lensing and the SKA observables. We use the values indicated in table \ref{tab:survey_details} for the different incarnations of the SKA surveys. Since we are considering cross-correlations with external data set, we need to restrict the sky fraction to the maximum common area between SKA surveys and Planck/SPT data. We therefore restrict the SKA surveys to 2500 sq. deq. when correlating with the SPT lensing map. For the correlation between  SKA surveys and the Planck lensing map we consider a maximum area of 15000 sq. deg. 

\begin{center}
\begin{figure}
\begin{minipage}[t]{0.48\linewidth}
\includegraphics[width=1\columnwidth]{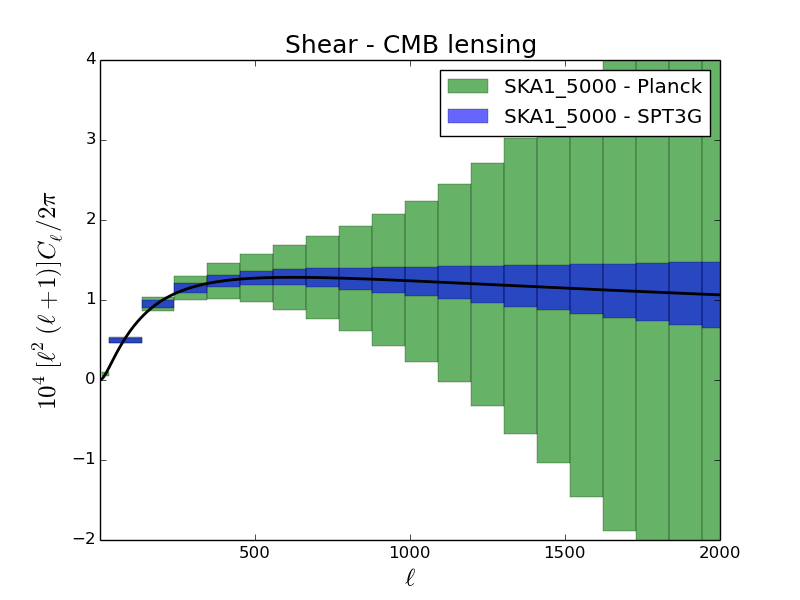}
\end{minipage}
\begin{minipage}[t]{0.48\linewidth}
\includegraphics[width=1\columnwidth]{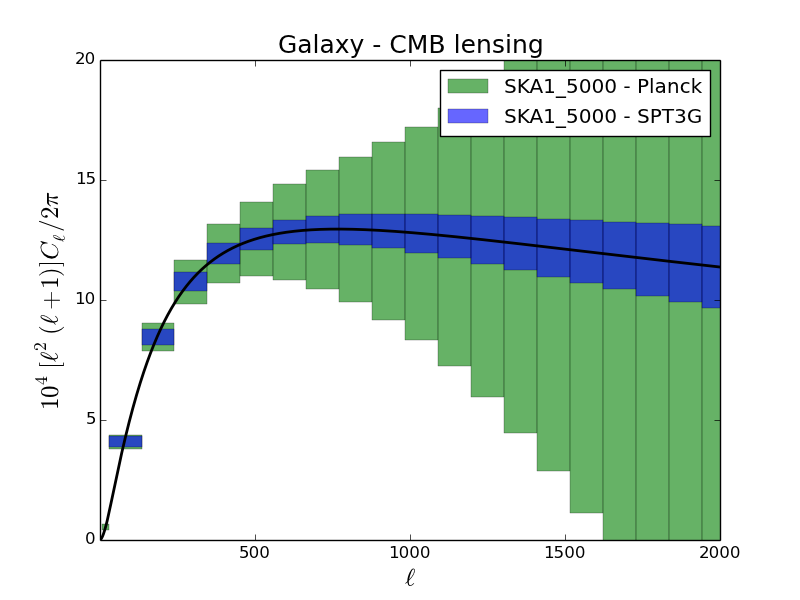}
\end{minipage}
\begin{minipage}[t]{0.48\linewidth}
\includegraphics[width=1\columnwidth]{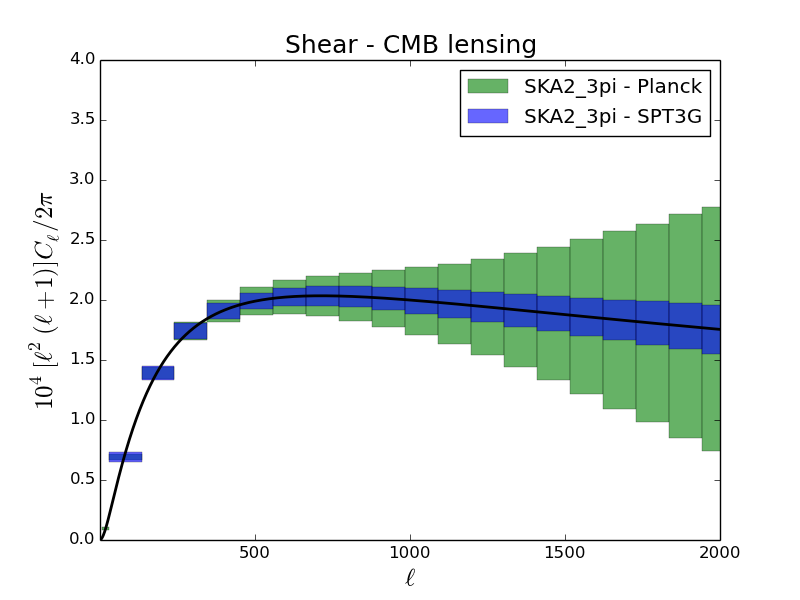}
\end{minipage}
\begin{minipage}[t]{0.48\linewidth}
\includegraphics[width=1\columnwidth]{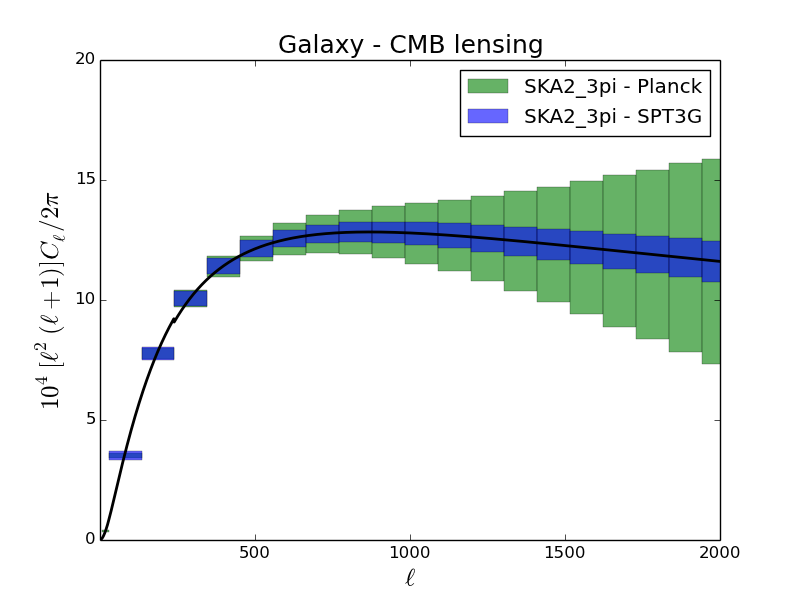}
\end{minipage}

\caption{ Weak-lensing (left column) and galaxy clustering (right column) for  SKA1-5000 (top row) and SKA2-$3\pi$ (bottom row) cross-correlations with the CMB lensing potential. Errors bars correspond to the Planck (green) and SPT3G (blue) noise levels.   \label{cmb_cor1}}
\end{figure}
\end{center}

Constraints on the amplitude of the cross-correlation between the SKA surveys observables (weak-lensing and galaxy clustering) and the CMB lensing potential are summarized in table \ref{tab:cmblens}. Some examples of those cross-correlations are shown in fig. \ref{cmb_cor1}.

\begin{table*}
   \centering
   \begin{tabular}{ |l|c|c c c c| } 
   \hline
Survey &  \shortstack{Area [deg$^2$] \\ Photo-z} & \multicolumn{4}{c|}{\shortstack{Planck$\;\;\;\;\;\;\;\;\;\;\;\; $  SPT3G \\ WGL  $\;\;$  \mbox{      } $\;$  LSS   \mbox{     } $\;$  WGL   \mbox{     } $\;\; $  LSS}} \\
\hline
SKA1 early & 5,000     	   & 5.7\% & 4.1\% &  2,8\% & 1.7\%  \\
SKA1       & 5,000     & 4.8\% & 3.4\% & 2.2\% & 1.3\% \\
SKA1       & $3\pi$     & 3.8\% & 2.2\% & 4.3\% & 1.9\% \\
SKA2       & 5,000  	  & 2.9\% & 2.4\% & 0.9\% & 0.7\% \\
SKA2       & $3 \pi$      & 1.6\% & 1.3\% & 1.1\% &0.9\%   \\
\hline
   \end{tabular}
   \caption{Constraints on the amplitude of the cross-correlation between the Planck and SPT3G lensing potential and the SKA surveys observables: weak lensing (WGL) and galaxy clustering (LSS).}
   \label{tab:cmblens}
\end{table*}

Prospects for cross-correlations between SKA and CMB lensing are very encouraging. As can be seen from the plots in fig. \ref{cmb_cor1}, most of the signal will come from the use of high resolution, small scale CMB data such as SPT3G (similar results would hold using the Atacama Cosmology Telescope (ACT) specifications). This can be seen as the blue errors bars being much smaller than the green ones, which represent the correlation with Planck. However, thanks to its large sky coverage, Planck will dominate on the largest scale, despite being intrinsically noisier than arcminute scale CMB data. 

The cross-correlation of SKA probes with CMB-lensing will improve our control of systematics and our ability to measure cosmology. CMB-lensing is sensitive to the integrated matter density between the observer and last-scattering without the mediation of biased tracers, helping us to control galaxy bias in a similar way to WGL but because the measurement of CMB-lensing is independent it can also calibrate important WGL systematics including shear measurement bias and intrinsic alignments, see Vallinotto (2013) for an example of the kinds of improvements that are possible.

\section{Conclusions.}

In this chapter we have investigated the synergies from cross-correlating different SKA datasets as well as cross-correlation with CMB lensing datasets from other missions. We conclude that, using internal cross-correlation, the SKA will be able to calibrate the redshift inaccuracies present in its weak lensing sample, optimising the possible statistical measurements of cosmology. We see from our results that LSS measurements with galaxies from the SKA are not very competitive during phase 1 but they are still important as they help with the aforementioned redshift calibration, particularly the planned $\sim$100 deg$^2$ deep survey which will overlap fully in redshift with the WGL survey source population. We estimate this calibration to be possible at the sub-per cent level with SKA2. 

We have shown that these cross-correlations provide huge gains for our Dark Energy analysis with SKA phase 2, but the largest gains come when we study Modified Gravity where, for SKA1 and SKA2, gains of several orders of magnitude are possible from combining WGL and LSS datasets. 

We also present constraints on the amplitude of the cross correlations between the lensing signal of future CMB experiments and the SKA and a sub present level constraint on the amplitude is possible. This would impose stringent constraints on the bias of SKA galaxy samples, calibrate systematic uncertainties in our WGL measurement and help constrain other cosmological parameters.

\bibliographystyle{apj}

\end{document}